\documentclass[12pt]{iopart}
\usepackage{iopams}
\usepackage{graphicx}

\usepackage{color}

\begin{document}

\title[]{Parity dependent Josephson current through a helical Luttinger liquid}

\author{S Barbarino$^1$, R Fazio$^1$, M Sassetti$^2$ and F Taddei$^1$}
\address{$^1$ NEST, Scuola Normale Superiore and Istituto Nanoscienze-CNR, I-56126 Pisa, Italy}
\address{$^2$ Dipartimento di Fisica, Universit\`{a} di Genova and CNR-SPIN, I-14146 Genova, Italy}
\ead{simone.barbarino@sns.it}

\begin{abstract}
We consider a superconductor-two dimensional topological insulator-superconductor junction (S-2DTI-S) and  study how the $2\pi$- and $4\pi$-periodic 
Josephson currents are affected by the electron-electron interaction. In the long-junction limit the  supercurrent can by evaluated by modeling the system  
as a helical Luttinger liquid coupled to superconducting reservoirs. After having introduced  bosonization in the presence of the parity constraint we turn 
to consider the limit of perfect and poor interfaces. For transparent interfaces, where perfect Andreev reflections occur at the boundaries, 
the Josephson current is marginally affected by the interaction. On the contrary, if strong magnetic scatterers are present in the weak link, the situation 
changes dramatically. Here Coulomb interaction plays a crucial role both in low and high temperature regimes. Furthermore, a phase-shift of Josephson 
current can be induced by changing the direction of the magnetization of the impurity.
\end{abstract}

\submitto{\NJP}

\maketitle

\tableofcontents

\section{Introduction}
Two-dimensional topological insulators (2DTI) are characterized by a gapped bulk spectrum and gapless edge states which are robust against 
time-reversal invariant perturbations~\cite{hasan2010,qi2011}. Originally predicted for HgTe/CdTe quantum wells in Refs.~\cite{Bernevig, Bernevig1}, 
and observed in Ref.~\cite{Konig}, these systems have attracted great attention in the last few years because of their peculiar electronic transport 
properties. Edge states possess a helical nature, namely electrons have spin direction and momentum locked to each other and constitute Kramer partners. 

Central for the present paper is the study of hybrid 2DTI-Superconductor (S) systems, a topic which has lately gained an increasing interest both 
theoretically  and experimentally. A comprehensive description of the activity in this field can be found in Ref.~\cite{Evelina}. The proximity effect into a 
2DTI has been largely investigated (see for example~\cite{Sato, Degiovanni,Recher,Alicea}). On the experimental side Andreev reflection at S-2DTI 
interfaces has been recently observed by Du in Ref.~\cite{Du}.  The, not yet observed, Josephson effect through a topological insulator is expected 
to show spectacular features. Indeed in 2009 Fu and Kane~\cite{FuKane} considered a Josephson junction with two s-wave superconductors 
connected by a weak link of length $L$, obtained  by a single edge of a 2DTI. In the short-junction regime (i.e. when $L\ll\xi$, with $\xi$ the BCS 
coherence length), they showed that the S-2DTI-S junction exhibits a fractional Josephson effect~\cite{Yakovenko,Kitaev,DasSarma,Jiang} in which 
the current phase relation has a $4\pi$-periodicity, rather than the standard $2\pi$-periodicity, if the fermion parity (parity of the number of electrons 
in the system) is preserved. This phenomenon is related to the presence of Majorana fermions~\cite{Kitaev} at the S-2DTI interfaces. 
More recently Beenakker \etal.~\cite{Beenakker} addressed the long-junction regime  ($L\gg\xi$) showing that the amplitude of the $4\pi$-periodic 
critical current, at zero temperature, is doubled with respect to the $2\pi$-periodic one. The AC Josephson effect in S-2DTI-S has also been 
considered~\cite{Meyer}. 

In all papers mentioned above, the effect of  the electron-electron interaction on the parity dependent Josephson current (JC) was neglected.
Aim of this paper is to study the $2\pi$- and $4\pi$-periodic Josephson effect in the long-junction regime for a S-2DTI-S system taking into 
account the Coulomb interaction within the framework of the helical Luttinger liquids~\cite{Wu,Teo}. Firstly we consider the case 
of transparent S-2DTI interfaces and then we address the presence of magnetic impurities in the weak link~\cite{Qi, Meng, Timm, Dolcetto}. 
Non-magnetic impurities cannot induce elastic back scattering in a helical liquid~\cite{footnote}.  We present analytical results for both high and low 
temperature regimes. If no impurities are present, at low temperature the JC exhibits a saw-tooth behavior, the $4\pi$-periodic critical current is 
doubled with respect to the $2\pi$-periodic one as in the non-interacting case. At high temperatures both the $2\pi$- and $4\pi$-periodic currents 
are sinusoidal and the $2\pi$-periodic current is suppressed with respect to the $4\pi$-periodic one. Our results agree with a recent paper by 
Cr\'epin \etal.~\cite{Trauzettel}. If point-like magnetic impurities are present within the weak link  the situation changes significantly. A single impurity 
with magnetization along the $z$-direction, which is the spin quantization axis of the helical states, induces a phase shift of the JC with respect 
to the transparent regime. Both the $2\pi$- and $4\pi$-periodic critical currents are not affected by such barrier. Otherwise, when the impurity 
magnetization lies in an arbitrary direction of the $xy$-plane, the current phase relation is always sinusoidal and the critical current depends 
on fermion parity and on the Luttinger interaction parameter. The previous results can be generalized if two barriers are present at the 2DTI-S interfaces.

The paper is organized as follows. In Section \ref{section1} we give a brief introduction to the bosonization technique with Andreev boundary conditions and 
discuss how the fermion parity can be implemented within the bosonization formalism. In Section \ref{section2} we discuss the $2\pi$- and $4\pi$-periodic
 JC for different regimes of the system. In Section \ref{section4} we summarize our results together with the conclusions of our work.

\section{Model of the S-2DTI-S system}
\label{section1}

\begin{figure}
	\centering
	\includegraphics[width=1\textwidth,trim=120pt 410pt 178pt 310pt, clip]{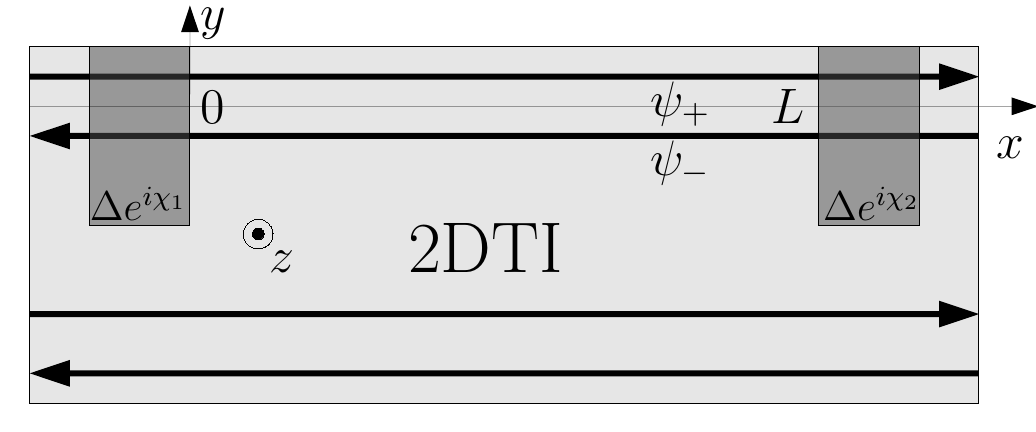}
	\caption{Helical states of a single edge of a two-dimensional topological 
	insulator sandwiched between two s-wave superconductors (shaded areas).}
	\label{A}
\end{figure}

The setup we consider is depicted in Figure~\ref{A}.  A single edge state of a 2DTI  is sandwiched between two s-wave superconducting terminals. 
The pair potential profile is 
\begin{equation}
	\Delta(x)=\Delta e^{i\chi_1} \mathcal{H}(-x)+\Delta e^{i\chi_2} \mathcal{H}(x-L)
\end{equation}
with  $\chi_1$, $\chi_2$  the macroscopic phases of the superconductors, $\Delta$  the modulus of the superconductive gap, and $L$ the distance 
between the two superconductors. We ignore self-consistency in determining the order parameter this is why we can safely  model the space profile of 
$\Delta$ using the step function $\mathcal{H}(x)$.  

The free Hamiltonian  of the edge of the  topological insulator is ($\hbar=1$):
\begin{equation}
	H_F=-i v_F\int_{0}^{L} \;dx\ \left(\psi^\dagger_+ \partial_x \psi_+ - \psi^\dagger_- \partial_x \psi_-\right) \; .
\end{equation}
The operator $\psi_{+/-}$ annihilates right/left moving spin up/down electrons and $v_F$ is the Fermi velocity. 
Short-range interactions between two electrons in the weak link can be analyzed in the so called $g$-ology~\cite{Giamarchi, Miranda} framework. 
In a helical Luttinger liquid, one has a forward scattering term 
$H_{S4}=g_4/2 \int_{0}^{L} \;dx\ \left(\psi^\dagger_{+}\psi_{+}\psi^\dagger_{+}\psi_{+}+ \psi^\dagger_{-}\psi_{-}\psi^\dagger_{-}\psi_{-}\right)$ and
a dispersive scattering term $H_{S2}=g_2 \int_{0}^{L} \;dx\ \psi^\dagger_{+}\psi_{+}\psi^\dagger_{-}\psi_{-}$. We have here neglected umklapp terms 
which are not relevant in the renormalization-group sense~\cite{Wu} if interactions are not too strong.

Point-like magnetic barriers in a generic point $0\leq \overline{x} \leq L$ of the weak link are described by the hamiltonian
\begin{equation}
	H_M = \Psi(\overline{x}) \vec{M} \cdot \vec{\sigma} \Psi^\dagger(\overline{x}) \;, 
\label{barriere}
\end{equation}
where $\vec{M}=(M_x,M_y,M_z)$ is the magnetization vector and $\vec{\sigma}= (\sigma_x,\sigma_y,\sigma_z)$ are the Pauli matrices acting on the 
spinor space $\Psi= \left( \begin{matrix} e^{ik_Fx} \psi_+, e^{-ik_Fx} \psi_-\end{matrix}\right)$, with $k_F$ the Fermi momentum. By expanding the scalar 
product in Eq. (\ref{barriere}), one obtains three terms: $H_{Mx}= M_x\left(e^{-2ik_Fx}\psi^\dagger_+\psi_- + e^{2ik_Fx} \psi^\dagger_- 
\psi_+ \right)$, $H_{My}= iM_y\left(e^{-2ik_Fx} \psi^\dagger_+ \psi_- -e^{2ik_Fx} \psi^\dagger_- \psi_+ \right)$ and $H_{Mz}= M_z\left(\psi^\dagger_+ 
\psi_+ - \psi^\dagger_- \psi_- \right)$. 

Finally the coupling of the edge to the superconducting electrodes, in the limit in which the superconducting gap is the largest energy scale in the 
problem, can be introduced through simple boundary conditions for the edge fermion field. Electrons  impinging at normal-superconductor  (in this case 
the normal part is the edge of the topological insulator) interface are retro-reflected as holes, this is the well known Andreev reflection.  By solving the 
Bogoliubov - de Gennes equations, one can find the boundary conditions for fermionic operators  which are essentially determined by Andreev reflections. 
In the limit $\Delta \rightarrow +\infty$, fermionic boundary conditions take a simple form because they are independent on the energy of the excitations 
(electrons can only be Andreev reflected and normal reflections do not occur), actually one finds~\cite{Loss} (at the two interfaces): 
\begin{eqnarray}
	\psi_+(x=0) &= &-ie^{i\chi_1} \psi^\dagger_- (x=0)\\
	\psi_+(x=L) &= &+ie^{i\chi_2} \psi^\dagger_- (x=L) \; . 
\end{eqnarray}
As shown in Ref.~\cite{Loss} such conditions, known as Andreev boundary conditions, are equivalent to twisted periodic boundary conditions for $\psi_-$ 
on an interval of length twice the length of the original system
\begin{equation}
	\psi_-(x+2L)=e^{i(\pi +\chi_2-\chi_1)} \psi_-(x) 
\label{2}
\end{equation}
supplemented by the connection between $\psi_+$ and $\psi_-$ following from the chiral symmetry
\begin{equation}
	\psi_+(x)=-ie^{i\chi_1}\psi^\dagger_-(-x)  \; . 
\label{1}
\end{equation} 
Boundary conditions~(\ref{2}) and~(\ref{1})  can be conveniently done in the bosonization language which we introduce in the following sub-section.

\subsection{Bosonization}
Fermionic operators can be put into the form $\psi_{\pm}(x)=\exp{[\pm i  \Phi_{\pm}(x)]}/\sqrt{2\pi a} $ in terms of the bosonic fields $ \Phi_{\pm}(x)$~\cite{Giamarchi}. 
The Andreev boundary conditions~(\ref{2}) and~(\ref{1}) are automatically satisfied if the boson fields are chosen to be~\cite{Loss}
\begin{equation}
	\cases{
	\Phi_+(x)=\varphi-\left(N+\frac{\chi}{\pi} \right)\frac{\pi x}{2L}+\rho(-x)\\
	\Phi_-(x)=\varphi+\left(N+\frac{\chi}{\pi} \right)\frac{\pi x}{2L}+\rho(x) 
\label{3}
} 
\end{equation}
with 
\begin{equation}
	\rho(x)=i \sum_{q>0}  \sqrt{\frac{\pi}{qL}}e^{-\frac{aq}{2}} \left(e^{-iqx} a^\dagger_q - e^{iqx} a_q\right) \;.
\end{equation}
In the previous equations $a \rightarrow 0^+$ is a convergence factor for the theory, $\chi \equiv \chi_2-\chi_1$; $\varphi$ and $N$ are conjugated zero-mode 
operators and the bosonic operators satisfy $[a_q,a^\dagger_{q'}]=\delta_{q,q'}$ with $q=\pi n/L$, $n\in \mathbb{Z}$.
Note that $\Phi_+$ and $\Phi_-$ obey canonical commutation relations and $[N,\varphi]=2i$, the eigenvalues of $N$ are even,  $N=2k$, $k \in \mathbb{Z}$ 
as implied by the boundary condition (\ref{2}).

It is convenient to define the bosonic fields $\Phi= \left(\Phi_-+\Phi_+\right)/2$ and  $\Theta=\left(\Phi_- -\Phi_+\right)/2$, thus fermion operators take the form 
$\psi_{\pm}(x)=\exp{[- i \Theta(x)\pm i\Phi(x)]}/\sqrt{2\pi a} $ where
\begin{equation}
	\hspace{-1.8cm}
	\cases{
	\Phi(x)=\varphi+i \sum_{q>0} \sqrt{\frac{\pi}{qL}}e^{-\frac{aq}{2}} \cos qx (a^\dagger_q - a_q) \equiv \varphi+\phi(x) \\
	\Theta(x)= \left(N+\frac{\chi}{\pi} \right)\frac{\pi x}{2L}+\sum_{q>0} \sqrt{\frac{\pi}{qL}}e^{-\frac{a q}{2}} \sin qx (a^\dagger_q+ a_q)\equiv 
	\left(N+\frac{\chi}{\pi} \right)\frac{\pi x}{2L}+\theta(x) . 
\label{op}}
\end{equation}

In terms of the bosonic  operators (\ref{op}) the Hamiltonian of the system $H=H_F+H_{S2}+H_{S4}$  (the superconducting electrodes enter only via the boundary conditions)
is the sum of a zero mode term $H_0$ and of a bosonic term $H_{B}$
\begin{equation}
	H \equiv H_0+H_{B}= \frac{u g \pi}{8L} \left(N+\frac{\chi}{\pi}\right)^2 + u \sum_{q>0} q a^\dagger_q a_q, 
\label{4}
\end{equation}
here $u=(1/2\pi) \left((2\pi v_F+g_4)^2 - g_2^2\right)^{\frac{1}{2}}$ is the renormalized Fermi velocity  and $g =((2\pi v_F +g_4-g_2)/(2\pi v_F +g_4+g_2))^{\frac{1}{2}}$
is the  Luttinger parameter which is related to the attractive ($g>1$) or repulsive ($g<1$) nature of the interaction.

In the bosonized form the Hamiltonian (\ref{barriere}) associated to the scattering from the magnetic impurities takes the form:
\begin{equation}
	H_{Mx} = \frac{M_x}{\pi a} \cos (\varphi + \phi_{\overline{x}}+k_F \overline{x}) 
\label{8}
\end{equation}
\begin{equation}
	H_{My}=\frac{M_y}{\pi a} \sin (\varphi + \phi_{\overline{x}}+k_F \overline{x}) 
\label{9}
\end{equation}
\begin{equation}
	H_{Mz}= M_z \partial_x \theta|_{x=\overline{x}} 
\label{10}.
\end{equation}
We have defined $\phi_{\overline{x}} \equiv \phi(\overline{x})$.

\subsection{Fermion parity}
In order to see how fermion parity is implemented in the bosonization language let's start by 
noting that the superconductive phase difference $\chi$ is defined up to multiples of $2m\pi$ with $m \in \mathbb{Z}$.  Consequently 
if we substitute $\chi \rightarrow \chi + 2n\pi$ ($n \in \mathbb{Z}$) in $\Phi_+$ and $\chi \rightarrow \chi + 2m\pi$ ($m \in \mathbb{Z}$) 
in $\Phi_-$, Eqs. (\ref{3}) become
\begin{equation}
	\cases{
	\Phi_+(x)=\varphi-\left(N+\frac{\chi}{\pi} \right)\frac{\pi x}{2L}+\rho(-x)-\frac{n \pi x}{L}\\
	\Phi_-(x)=\varphi+\left(N+\frac{\chi}{\pi} \right)\frac{\pi x}{2L}+\rho(x)+\frac{m \pi x}{L}.
	}
\label{fiffi}
\end{equation}
Using Eq. (\ref{1}) with $\chi_1$ set to zero through a proper gauge transformation, we get $n=-m$ in Eqs. (\ref{fiffi}). We obtain the periodicity requirements
\begin{equation}
	\cases{
	\Phi_+(x+2L)= \Phi_+(x) - \pi \left(N+\frac{\chi}{\pi} \right)+2m\pi \\
	\Phi_-(x+2L)= \Phi_-(x) + \pi \left(N+\frac{\chi}{\pi} \right)+2m\pi.
	} 
\label{ericee}
\end{equation}
satisfied by the even eigenvalues of the operator $N$, $N=2k$, $k \in \mathbb{Z}$. The field $\Phi(x)$ corresponding to the periodicity requirements (\ref{ericee}) 
which obeys the condition $\Phi(x+2L)= \Phi(x)+2m\pi$,  will be called parity independent in the rest of the paper. We now assume that the superconductive 
phase difference $\chi$ is defined up to multiples of $4m\pi$, instead of $2m\pi$, with $m \in \mathbb{Z}$.  By carrying out the same procedure  used for the 
parity independent case we obtain the analogous of Eqs. (\ref{ericee}):
\begin{equation}
\cases{
	\Phi_+(x+2L)= \Phi_+(x) - \pi \left(N+\frac{\chi}{\pi} \right)+4m\pi \\
	\Phi_-(x+2L)= \Phi_-(x) + \pi \left(N+\frac{\chi}{\pi} \right)+4m\pi.
	} 
\label{ericeee}
\end{equation}
that lead to $\Phi^{(E)}(x+2L)=\Phi^{(E)}(x)+4m\pi$, which will be called even parity dependent~\cite{Fisher}. Periodicity requirements given in Eqs. (\ref{ericeee}) 
are satisfied if and only if the eigenvalues of the operator $N$ take the form $N=4k$, $k \in \mathbb{Z}$. Furthermore, by inducing an additional shift of the 
superconductive phase $\chi$ of $2\pi$ in Eqs. (\ref{ericeee}) we obtain the conditions 
\begin{equation}
	\cases{
	\Phi_+(x+2L)= \Phi_+(x) - \pi \left(N+\frac{\chi}{\pi} \right)+(4m+2)\pi \\
	\Phi_-(x+2L)= \Phi_-(x) + \pi \left(N+\frac{\chi}{\pi} \right)+(4m+2)\pi 
	} 
\label{ericeeee}
\end{equation}
that lead to $\Phi^{(O)}(x+2L)=\Phi^{(O)}(x)+(4m+2)\pi$, which will be called odd parity dependent~\cite{Fisher}.  Periodicity requirements (\ref{ericeeee}) 
are now satisfied if and only if the eigenvalues of the operator $N$ take the form $N=4k+2$, $k \in \mathbb{Z}$.

We can reach the same conclusions  by studying the spectrum of the Hamiltonian (\ref{4}) which is unchanged by a proper shift of the superconductive phase $\chi$. 
If fermion parity is not conserved, namely the hamiltonian is invariant with respect to a shift of the form $\chi \rightarrow \chi + 2\pi m$, one obtains that 
$N \rightarrow N-2m$ and concludes the eigenvalues of $N$ must be of the form $N=2k$. If fermion parity is conserved $\chi \rightarrow \chi + 4\pi m$, one 
obtains the constraint $N \rightarrow N-4m$ or $N \rightarrow N-(4m+2)$ and concludes that the eigenvalues must be of the form $N=4k$ in the even case 
and $N=4k+2$ in the odd one. Incidentally note that the number of fermions in the weak link is $N/2$~\cite{vonDelft}.
It is important to stress  that constraints imposed by fermion parity conservation involve only the eigenvalues of the zero mode operator $N$ and leave 
the bosonic excitation modes unaffected. 
Summarizing, the possible eigenvalues of the operator $N$ are $N=2k$, $k \in \mathbb{Z}$ if parity is not conserved and $N=4k$ (even) or $N=4k+2$ 
(odd), $k \in \mathbb{Z}$ in the opposite case.

\section{The Josephson current} 
\label{section2} 
Equipped with the definitions given above we calculate the $2\pi$- and $4\pi$-periodic JC for different configurations of the system. 
Firstly we focus on the transparent regime where perfect Andreev reflections occur in correspondence of the S-2DTI interfaces, then we introduce 
magnetic barriers which induce normal reflections with spin-flip. The Josephson current can be computed from the partition function $Z(\chi)$  as
\begin{equation}
	I_J(\chi)=-\frac{2e}{\beta} \frac{\partial \ln Z(\chi)}{\partial \chi} \label{5}
\end{equation}
where $e$ is the elementary charge, $\beta=1/T$, $T$ the temperature and the Boltzmann constant $k_B=1$.  The partition function can be expressed 
in the form
\begin{equation}
	Z(\chi)=\sum_{m=-\infty}^{\infty} \int_{\Delta \varphi = 2\pi m} \mathcal{D}\varphi  \int \prod_{q>0} 
	\mathcal{D}a^\dagger_q \mathcal{D}a_q \exp{[S_0+S_{B}+S_M]}
\label{7}
\end{equation}
where $\Delta \varphi = \varphi(\beta)-\varphi(0)=2\pi m$ and $m \in \mathbb{Z}$; $S_M$ is the Euclidean action corresponding to the Lagrangian of magnetic barriers, $S_{0}$ 
and $S_{B}$ to the Lagrangian of the Luttinger Hamiltonian introduced in Eq. (\ref{4}):
\begin{equation}
L_{B}= -\sum_{q>0} a^\dagger_q \partial_\tau a_q - u\sum_{q>0} q a^\dagger_q a_q,
\end{equation}
\begin{equation}
L_0=\frac{\partial H_0}{\partial n} n -H_0= i \frac{\chi}{\alpha \pi} \partial_\tau \varphi - \frac{2L}{\alpha^2 \pi ug} (\partial_\tau \varphi)^2 ,\label{erice}
\label{6}
\end{equation}
where
\begin{equation}
\partial_\tau \varphi = i \frac{\partial H_0}{\partial n} = i\frac{\alpha u g \pi}{4L} \left(\alpha n + \frac{\chi}{\pi}\right). \label{erice2}
\end{equation}
Here $n\equiv N/\alpha$, $n \in \mathbb{Z}$ and $\alpha=2$ if the fermion parity is not conserved, i.e. $2\pi$-periodic case, and $\alpha=4$ 
if the fermion parity is preserved, i.e. $4\pi$-periodic even case.  In the next section we will also show how to calculate the $4\pi$-periodic odd 
current. Moreover, we stress again that constraints on the eigenvalues of the zero mode operator $N$ due to the fermion parity symmetry 
affect the Lagrangian $L_0$, as one can see from Eqs. (\ref{erice}, \ref{erice2}), but not $L_{B}$.

\subsection{Transparent interfaces}
\label{4maggio}
Firstly we calculate  the $2\pi$- and $4\pi$-periodic supercurrent in the transparent regime for both low and high temperatures.  The evaluation of the 
partition function is straightforward  because Eq. (\ref{7}) can be put into the form $Z(\chi)=Z_0(\chi) Z_{B}$, with
\begin{equation}
	\hspace{-1cm}
	Z_0(\chi)=\sum_{m=-\infty}^{\infty} \int_{\Delta \varphi = 2\pi m} \mathcal{D}\varphi \exp{\left[i \frac{\chi}{\alpha \pi} \int_{0}^{\beta} d\tau 
 	\partial_\tau \varphi - \frac{2L}{\alpha^2 \pi ug} \int_{0}^{\beta} d\tau(\partial_\tau \varphi)^2\right]}.
\label{erice3}
\end{equation} 
$Z_{B}$ is independent on $\chi$ and does not contribute to the Josephson current, as one can easily verify from Eq. (\ref{5}).
In order to perform the path integral in Eq. (\ref{erice3}), we parametrize $\varphi(\tau)$ as  $\varphi(\tau) = 2\pi m \tau/\beta + \tilde{\varphi} (\tau)$,  
with $\tilde{\varphi}(0)=\tilde{\varphi}(\beta)$ obtaining 
\begin{equation}
	Z_0(\chi)=\sum_{m=-\infty}^{\infty} e^{i \frac{2m\chi}{\alpha} -\frac{8\pi m^2 L}{\alpha^2 ug \beta}} 
	\int_{\tilde{\varphi}(0)=\tilde{\varphi}(\beta)} \mathcal{D}\tilde{\varphi} \exp{\left[-\frac{L}{\alpha^2 \pi ug} 
	\int_0^ \beta d\tau \left(\partial_\tau \tilde{\varphi} \right)^2\right]} \; .
\label{erice4}
\end{equation}
The integral in Eq. (\ref{erice4}) does not contribute to the supercurrent since it is independent on $\chi$. Using Poisson's summation formula~\cite{Gradstein} and neglecting constants which do not contribute to the JC, one has
\begin{equation}
	Z_0(\chi) \propto \sum_{m=-\infty}^{\infty} e^{-\frac{\beta ug \pi}{8L} \left( \alpha m +
	\frac{\chi}{\pi} \right)^2}= \frac{2}{\alpha} \sqrt{\frac{2}{Ag}} \theta_3 \left(\frac{\chi}{\alpha}, e^{-\frac{8\pi}{A\alpha^2 g}}\right),
\end{equation}
$\theta_3$ is the elliptic Jacobi's function and  $A \equiv \beta u/L$.  In the low temperature regime $A\gg1$, the supercurrent exhibits a saw-tooth behavior
\begin{equation}
	I_J (\chi)= \frac{e ug}{L} \frac{\chi}{2\pi} \hspace{1cm}; \hspace{1cm} |\chi| < \frac{\alpha}{2} \pi \; .
\label{4maggio2}
\end{equation}
As shown by Beenakker \etal.~\cite{Beenakker} in the non-interacting case, the $4\pi$-periodic critical current is twice the $2\pi$-periodic one even 
in the presence of interactions. In the high temperature regime $A\ll 1$, the JC
\begin{equation}
	I_J(\chi)= \frac{8e}{\alpha \beta} e^{-\frac{8\pi}{A\alpha^2 g}} \sin \frac{2\chi}{\alpha} 
\label{4maggio3}
\end{equation}
has a sinusoidal behavior and the ratio between the $4\pi$-periodic critical current ($\alpha=4$) and the $2\pi$-periodic critical one ($\alpha=2$) is much 
larger than 2, since $A\ll 1$.

Let's now focus on the role played by interactions. For a generic interaction with $g_2 \neq g_4$, one has $ug= v_F+(g_4-g_2)/(2\pi)$, i.e. 
the forward scattering term and the dispersive one act differently on the critical valued of the JC. Conversely, for the Coulomb interaction, one has
$g_2=g_4$, as shown in~\cite{Flensberg}. In this case, the JC is completely unaffected by the Coulomb interaction because $ug=v_F$. 

In the odd parity conserving case the partition function has the form
\begin{equation}
	Z_0(\chi)\propto \left.\sum_{m=-\infty}^{\infty} e^{-\frac{\beta ug \pi}{8L} \left( \alpha m +2+\frac{\chi}{\pi} \right)^2}\right|_{\alpha=4}= 
	\left.\frac{2}{\alpha} \sqrt{\frac{2}{Ag}} \theta_3 \left(\frac{\chi+2\pi}{\alpha}, e^{-\frac{8\pi}{A\alpha^2 g}}\right)\right|_{\alpha=4}
\end{equation}
from which one can easily see that the corresponding current is equal to the even one translated of $2\pi$.

\subsection{One impurity}
We start by considering a single magnetic impurity described by the Hamiltonian (\ref{10}) in a generic point of the weak link whose magnetization has the same direction 
of the spin quantization axis, namely $z$-axis. The calculation proceeds similarly to the transparent regime, where $L_0$ is now given by 
\begin{equation}
	L_0=\frac{\partial H_0}{\partial n} n -H_0= i \left(\frac{\chi}{\alpha \pi}+\frac{2M_z}{\alpha ug} \right) \
	\partial_\tau \varphi - \frac{2L}{\alpha^2 \pi ug} (\partial_\tau \varphi)^2
\end{equation}
with
\begin{equation}
	\partial_\tau \varphi = i \frac{\partial H_0}{\partial n} = i\frac{\alpha u g \pi}{4L} \left(\alpha n + \frac{\chi}{\pi}\right) +i \frac{\pi \alpha}{2L}M_z.
\end{equation}
From the partition function
\begin{equation}
	Z_0(\chi)\propto \frac{2}{\alpha} \sqrt{\frac{2}{Ag}} \theta_3 \left(\frac{\chi}{\alpha}+\frac{2M_z \pi}{\alpha u g}, e^{-\frac{8\pi}{A\alpha^2 g}}\right),
\end{equation}
we get the Josephson current in the low temperature regime  
\begin{equation}
	I_J(\chi)= \frac{eug}{L} \frac{\chi}{2\pi}+\frac{e M_z}{L} \hspace{0.75cm};\hspace{0.75cm} \left|\chi + \frac{2\pi M_z}{ug}\right|<\frac{\alpha}{2} \pi 
\label{4maggio4}
\end{equation}
and in the high temperature regime
\begin{equation}
	I_J(\chi)= \frac{8e}{\alpha \beta} e^{-\frac{8\pi}{A\alpha^2 g}} \sin \left(\frac{2\chi}{\alpha}+\frac{4M_z \pi}{\alpha ug}\right). 
\label{4maggio5}
\end{equation}
The values of the critical currents remain unchanged with respect to the transparent regime found in Sub-section~\ref{4maggio}. The current-phase relation, 
however, exhibits a phase shift whose magnitude depends on $M_z$ and, for the high-temperature regime,  on the periodicity of the JC.  
Note that the supercurrent remains finite even at $\chi=0$ because time reversal symmetry is broken by the magnetic barrier.

We now consider a magnetic impurity in $x=0$, or equivalently in $x=L$, with magnetization lying in an arbitrary direction of the $xy$-plane. 
The corresponding Hamiltonian is the sum of the  Hamiltonians given in Eqs. (\ref{8},\ref{9}) 
\begin{equation}
	H_M = H_{Mx}+H_{My} = \frac{M_x}{\pi a} \cos (\varphi + \phi_0)+ \frac{M_y}{\pi a} \sin (\varphi + \phi_0)
\end{equation}
that can be conveniently written as
\begin{equation}
	H_{M} =  \frac{|M|}{\pi a} \cos(\varphi + \phi_0 + \delta_0) \label{12}
\end{equation}
with $|M|=\sqrt{M_x^2+M_y^2}$ and $\tan (\delta_0-\pi/2) =M_y/M_x$, $-\pi \leq \delta_0\leq \pi$; $\phi_0=\phi(x=0)$.
The impurity potential is a relevant term in the renormalization group sense~\cite{KaneFisher} if $g<1$, then we consider the strong barrier limit, where 
the argument of the $\cos$  function in Eq. (\ref{12}) is strongly pinned in the minima.
The partition function of the system is given by Eq. (\ref{7}), where $S_M =-\int_0^\beta d\tau H_{M}$ and $S_0$ consists of a linear term $S_{0,l}$ in 
$\partial_\tau \varphi$ and of a quadratic term $S_{0,q}$, as shown in Eq. (\ref{6}). Consequently $Z(\chi)$ can be expressed in the form
\begin{equation}
	Z(\chi) \propto Z_0  \left[ 1 + 2\sum_{m=1}^{\infty} \cos \left(\frac{2m}{\alpha} \chi \right) \frac{Z_m}{Z_0} \right] 
\label{14}
\end{equation}
where 
\begin{equation}
	Z_m= \int_{\Delta \varphi=2\pi m} \mathcal{D}\varphi \int \mathcal{D} \phi_0 \exp \left [S_{0,q}+S_{B}^{eff}+S^\Lambda+S_{M}\right].
\end{equation}
Here 
\begin{equation}
S_{B}^{eff} = -\frac{1}{2\pi g \beta}  \sum_{\omega} \frac{\omega^2}{-\frac{u}{L}+  \omega \coth \frac{L \omega}{u}} |\phi_{0}(\omega)|^2
\end{equation}
is an effective action obtained by integrating the degrees of freedom of $S_{B}$ away from the impurity (more details are given in \ref{App2} where 
the effective action is calculated for a generic point $\overline{x}$ in the weak link);  $\omega=2\pi n / \beta$, $n \in \mathbb{Z}$ are the  Matsubara frequencies and 
\begin{equation}
	S^\Lambda = -\frac{M_0}{2}  \int_0 ^\beta d\tau (\partial_\tau \phi_0)^2
\end{equation}
is a high-frequency cut-off action with $M_0 = 1/\Lambda \approx 1/\Delta$.
We evaluate the partition function in the semiclassical limit simply searching the stationary path of the action $S \equiv S_{0,q}+S^\Lambda+S_{M}$ 
which gives the most relevant contribution to the functional integral. Such procedure is justified by the strong magnetic barrier limit. Then, we 
include the contributions of $S_{B}^{eff}$, which plays the role of a dissipative environment, by integrating out the low energy fluctuations of 
$\varphi$ and $\phi_0$ around the stationary path~\cite{Takane, Takane1, Takane2, Furusaki}.

It is convenient to introduce the fields 
\begin{equation}
	\cases{
	\phi_r=\varphi + \phi_0\\
	\phi_R=\frac{1}{m_L+M_0} (m_L \varphi - M_0 \phi_0)
	}
\end{equation}
where $m_L=4L/(\alpha^2 \pi ug)$.  The action $S$ takes then the form
\begin{equation}
	S= - \int_0 ^{\beta} d\tau \left(\frac{M_r}{2} (\partial_\tau \phi_r)^2 + \frac{M_R}{2}(\partial_\tau \phi_R)^2 + 
	\frac{|M|}{\pi a} \cos (\phi_r+\delta_0)\right) 
\label{13maggio1}
\end{equation}
with $M_r= M_0 m_L/(m_L+M_0)$, $M_R= m_L+M_0$. The stationary requirement leads to
\begin{equation}
	\cases{
	\frac{\delta S}{\delta \phi_r}=-M_r \partial^2_\tau \phi_r + \frac{|M|}{\pi a} \sin (\phi_r+\delta_0)=0\\
	\frac{\delta S}{\delta \phi_R}=-M_R \partial^2_\tau \phi_R=0}
\end{equation}
with the boundary conditions $\phi_r(\beta)-\phi_r(0)=2\pi m$ and $\phi_R(\beta)-\phi_R(0)=2m\pi m_L/(m_L+M_0)$.
As $m_L\gg M_0$, one gets $\phi_R(\beta)-\phi_R(0)\approx 2\pi m$ and $\phi_R^{st} \approx \varphi^{st} = 2\pi m \tau/\beta + 2\pi l$, $l \in \mathbb{Z}$.
The saddle point solution for $\phi_r(\tau)$ obeys a Sine-Gordon equation which admits the instanton solution 
 \begin{equation}
	\phi_r^{st} (\tau)= 2\arccos \left[-\tanh\left(\sqrt{\frac{|M|}{\pi a M_0}} \tau\right) \right]+2\pi l_0 -\delta_0
\end{equation}
with $l_0 \in \mathbb{Z}$. The solution for $\phi^{st}_0(\tau)$ is
\begin{equation}
	\phi_0^{st}(\tau)=\sum_{j=m}^J e_j \phi_r^{st}(\tau-\tau_j)- \frac{2\pi \tau}{\beta}m - 2\pi l  
\label{13maggio}
\end{equation}
with $\sum_{j=m}^J e_j=m$,  $J \in \mathbb{N}$ is the number of instantons. By calculating the saddle point action~\cite{Schmid}, one can show 
that the quantity $Z_m/Z_0$ is proportional to  $\delta^J\ll 1$, where $\delta=\lambda^{-\frac{1}{2}} \exp{\left(-2M_r \sqrt{\lambda}\right)}$, 
with $\lambda \equiv|M|/(\pi a M_0)= \Lambda |M|/(\pi a)\gg 1$. In this limit, we can take $m= 1$ with $J=1$ and Eq. (\ref{13maggio}) can be 
approximated $\phi_0^{st}(\tau)=2\pi \mathcal{H}(\tau-\tau_1)-2\pi \tau/ \beta -\delta_0+2\pi (l_0-l)$ whose Fourier transform is
\begin{equation}
	\phi_0^{st}(\omega)=\frac{2i\pi}{\omega} e^{i\omega \tau_1}+(\pi(2\beta l+ \beta-2\tau_1)-\delta_0 \beta) \delta_{\omega,0}.
\end{equation}
with $l \equiv l_0-l$. Due to the strong impurity potential, $\varphi$ and $\phi_0$ can fluctuate under the condition that $\varphi + \phi_0$ is strongly 
pinned. Such low energy fluctuations can be taken into account by introducing the field $\psi$
\begin{equation}
	\cases{
	\varphi= \varphi^{st}+\psi\\
	\phi_0= \phi_0^{st}-\psi.
	}
\end{equation}
The partition function becomes
\begin{equation}
	\frac{Z_1}{Z_0}= \delta e^{-\frac{8\pi L}{\alpha^2 ug \beta}} \int_{-\frac{\beta}{2}}^{\frac{\beta}{2}} d\tau_1  \int \mathcal{D} \psi \exp{[-S_{ins}]} 
\label{13}
\end{equation}
with
\begin{equation} 
	S_{ins}= \frac{1}{2\pi g \beta}\sum_{\omega} \left( \frac{4L\omega^2}{\alpha^2 u}|\psi(\omega)|^2+\frac{\omega^2}{-\frac{u}{L}+
	\omega \coth \frac{L \omega}{u}}  |\phi_0^{st}(\omega)- \psi(\omega)|^2 \right) 
\label{16}
\end{equation}
and integration over $\psi$ gives
\begin{equation}
	\frac{Z_1}{Z_0}= \delta e^{-\frac{8\pi L}{\alpha^2 ug \beta}}  \int_{-\frac{\beta}{2}}^{\frac{\beta}{2}} d\tau_1 \exp{[-\tilde{S}_{ins}]} ,
\end{equation}
with
\begin{equation}
	\tilde{S}_{ins}= \frac{2\pi}{ g \beta} \sum_{\omega \neq 0 } \frac{1}{\epsilon \frac{u}{L}+\omega \coth \frac{L \omega}{u}}   \;. 
\end{equation}
In the previous expression $\epsilon = \alpha^2 /4-1$: if $\alpha=2$, $\epsilon=0$;  if $\alpha=4$, $\epsilon=3$; remarkably no contribution to $\tilde{S}_{ins}$ 
arises from the $\omega=0$ term. By using the definition (\ref{5}) of the JC definition one finds (more details are given in \ref{App1a}) for the high temperature regime
\begin{equation}
	I_J(\chi)=\delta \frac{8e}{\alpha}e^{-\frac{8\pi L}{\alpha^2 ug\beta}} \left(\frac{2\pi}{e^\gamma \Lambda \beta}\right)^{\frac{2}{g}} 
	\sin \frac{2\chi}{ \alpha}, 
\label{4maggio6}
\end{equation}
which is sinusoidal both in the $2\pi$- and $4\pi$-periodic case. Moreover the ratio between the $4\pi$-periodic critical current and the $2\pi$-periodic 
critical one is much larger than $1$. Keeping fixed the parity, the critical current is reduced by the Coulomb interaction according to the power $2/g$.
In the low temperature regime one obtains
\begin{equation}
	I_J(\chi)= \delta \frac{8e}{2} e^{-\frac{2}{g}(\gamma+2\ln 2)}  \left( \frac{\pi u } {\Lambda L} \right)^{\frac{2}{g}} \sin \chi, 
\label{4maggio7}
\end{equation}
for the $2\pi$-periodic JC and
\begin{equation}
	I_J(\chi)= \delta \frac{8e}{4} \left( \frac{\pi u } {\Lambda L} \right)^{\frac{2}{g}} \sin \frac{\chi}{2}, 
\label{4maggio8}
\end{equation}
for the $4\pi$-periodic JC (even). As expected, the odd $4\pi$-periodic current can be obtained by translating the even $4\pi$-periodic current of $2\pi$. 
 In the non-interacting case the $4\pi$-periodic current is larger than the $2\pi$-periodic one, as in the transparent regime. The sinusoidal behavior is a direct 
 consequence of the strong barrier limit and it is not related to helical nature of the weak link: Andreev reflections at S-2DTI interfaces are strongly suppressed 
 with respect to normal reflections induced by magnetic barriers.  Interestingly, if the repulsive interaction is strong, the $4\pi$-periodic critical current is 
 more robust with respect to the $2\pi$-periodic because it lacks of the exponential term $\exp[-2g^{-1}(\gamma+2\ln 2)] $, thus the ratio between the 
 $4\pi$-periodic JC and the $2\pi$-periodic one is bigger than $2$.  We note that the JC is unaffected by changing the direction of the magnetization 
 in the $xy$-plane because Eqs. (\ref{4maggio6}, \ref{4maggio7},\ref{4maggio8}) do not depend on $\delta_0$. 

Finally we point out that by varying the position of the barrier keeping fixed the direction of the magnetization in the $xy$-plane, different power laws are obtained. 
For example, if the barrier is in $L/2$ of the weak link, one obtains $4/g$ instead of $2/g$ for both high and low temperature regimes, i.e. in the 
middle of the junction the barrier  acts strongly than at the interfaces.

\subsection{Two impurities}
\label{Bxy2}
Finally we discuss the case of  two magnetic impurities at the S-2DTI interfaces, namely in $x=0$ and $x=L$. The two magnetizations, equal 
in magnitude, lying in the $xy$ plane and not collinear, are taken into account by the Hamiltonian
\begin{equation}
	H_M=\frac{|M|}{\pi a} \left[\cos (\varphi + \phi_0+\delta_1) +\cos (\varphi + \phi_L+\delta_2) \right] ,
\end{equation}
with $\phi_0\equiv \phi(0,\tau)$ and $\phi_L\equiv \phi(L,\tau)$, or equivalently 
\begin{equation}
	H_M=\frac{2|M|}{\pi a} \cos \left(\varphi + \overline{\phi}+\overline{\delta}\right) \cos \left(\frac{\tilde{\phi}}{2}+\frac{\tilde{\delta}}{2}\right) ,
\label{15}
\end{equation}
with $\overline{\phi}= 1/2\left(\phi_L + \phi_0\right)$, $\tilde{\phi}= \phi_L - \phi_0$. The parameters $\delta_1$ and $\delta_2$ specify the angle of 
magnetizations with respect to the $y$-axis, $\overline{\delta}\equiv(\delta_1+\delta_2)/2$  and $\tilde{\delta}\equiv \delta_2-\delta_1$, $k_FL$ has been 
chosen proportional to $\pi$. The partition function takes the  form  
\begin{equation}
	Z_m= \int_{\Delta \varphi=2\pi m} \mathcal{D}\varphi \int \mathcal{D} \overline{\phi} \int \mathcal{D} \tilde{\phi} \exp 
	\left [S_{0,q}+S_{B}^{eff}+S^\Lambda+S_M\right],
\end{equation}
$S_{B}^{eff}$ is the effective action obtained by integrating out the bosonic modes away from $x=0$ and $x=L$ of the Hamiltonian (\ref{15})~\cite{Takane2}.
\begin{equation}
	S_{B}^{eff}= -\frac{1}{\pi g\beta} \sum_\omega \overline{J}^{-1}(\omega) |\overline{\phi}(\omega)|^2-\frac{1}{4\pi g\beta} 
	\sum_\omega \tilde{J}^{-1}(\omega) |\tilde{\phi}(\omega)|^2
\end{equation}
with 
\begin{equation}
	\overline{J}(\omega)=\frac{1}{\omega} \coth \frac{L \omega}{2u} - \frac{2u}{L\omega^2} \hspace{1cm};\hspace{1cm}
	\tilde{J}(\omega)=\frac{1}{\omega} \tanh \frac{L\omega}{2u},
\end{equation}
and
\begin{equation}
S^\Lambda = - \int_0^{\beta} d\tau  \left (\frac{\overline{M}}{2} (\partial_\tau \overline{\phi})^2 + \frac{\tilde{M}}{2} (\partial_\tau \tilde{\phi})^2\right) 
\end{equation}
provides the high frequency cut-off ($\overline{M}= 2/\Lambda \approx 2/\Delta$ and $\tilde{M} = 1/(2\Lambda)=1/(2\Delta)$).

We proceed as in the single impurity problem and we calculate the functional integral in the semiclassical approximation.  By introducing the fields
\begin{equation}
	\cases{
	\phi_r = \varphi + \overline{\phi}\\
	\phi_R = \frac{1}{\overline{M}+m_L}\left(m_L \varphi -\overline{M} \hspace{0.1cm} \overline{\phi}\right)
	}
\end{equation}
with $m_L=4L/(\alpha^2\pi ug)$.  The saddle point for $S_{0,q}+S^\Lambda+S_M$ leads to the equations 
\begin{equation}
	\cases{
	\frac{\delta S_{st}}{\delta \phi_r}=-M_r \partial_\tau^2 \phi_r - \frac{2|M|}{\pi a} \sin \left(\phi_r+\overline{\delta}\right) 
	\cos \left(\frac{\tilde{\phi}}{2}+\frac{\tilde{\delta}}{2}\right) =0\\
	\frac{\delta S_{st}}{\delta \phi_R}=-M_R \partial_\tau^2 \phi_R=0\\
	\frac{\delta S_{st}}{\delta \tilde{\phi}}=-\tilde{M} \partial_\tau^2 \tilde{\phi} - \frac{|M|}{\pi a}\cos \left(\phi_r+\overline{\delta}\right) 
	\sin \left(\frac{\tilde{\phi}}{2}+\frac{\tilde{\delta}}{2}\right) =0
	}
\end{equation}
with the boundary conditions $\phi_R(\beta)=\phi_R(0)+2m\pi m_L/(m_L+\overline{M})$ and $\phi_r(\beta)= \phi_r(0)+2\pi m$, 
where $M_r= m_L \overline{M}/(m_L + \overline{M})$ and $M_R=m_L+\overline{M}$. 
In the strong barrier limit it is sufficient to consider only $m=1$. The instanton solutions take the form
\begin{equation}
	\cases{
	\overline{\phi}^{st}(\tau) = \pi \mathcal{H} (\tau - \tau_1) + \pi \mathcal{H}(\tau - \tau_2)+\pi(l_1+l_2)-\frac{2\pi \tau}{\beta}-2\pi l-\overline{\delta}\\
	\tilde{\phi}^{st} (\tau) = 2\pi  \mathcal{H}(\tau - \tau_1) - 2\pi  \mathcal{H}(\tau - \tau_2)+2\pi(l_2-l_1)  -\tilde{\delta}
	}
\end{equation}
whose Fourier transform are
\begin{equation}
	\cases{
	\overline{\phi}^{st}(\omega)= \frac{\pi i}{\omega} \left(e^{i\omega \tau_1}+e^{i\omega \tau_2}\right) +\left(\pi \overline{l} \beta+ 
	\pi \beta-\pi\tau_1-\pi\tau_2 -\overline{\delta} \beta\right)	\delta_{\omega,0}\\
	\tilde{\phi}^{st}(\omega)= \frac{2\pi i}{\omega} \left(e^{i\omega \tau_1}-e^{i\omega \tau_2}\right)+
	(2\pi \tilde{l}\beta +2\pi\tau_2-2\pi\tau_1-\tilde{\delta}\beta)\delta_{\omega,0}
	}
\end{equation}
with $\tilde{l}=l_1+l_2-2l $ and $\overline{l}=l_1-l_2$.
We substitute  $\overline{J}(\omega)$ and $\tilde{J}(\omega)$ in the action $S_{0,q}+S_{B}^{eff}$ and we take into account the
low energy fluctuations  of $\varphi + \overline{\phi}$ as in the single impurity problem by introducing the auxiliary fields $\psi$, 
while $\tilde{\phi}$ is kept strongly pinned. Integration over $\psi$ gives two contributions
\begin{equation}
	\tilde{S}_{ins,\omega}=  \frac{2\pi}{g\beta} \sum_{\omega \neq 0} \left(\frac{2}{\epsilon \frac{u}{L} + \omega 
	\coth \frac{L \omega}{2u}}+\frac{\epsilon \frac{u}{L}+\frac{2\omega}{\sinh{\frac{L	\omega}{u}}}}{\omega^2+
	\epsilon \omega \frac{u}{L} \tanh{\frac{L \omega}{2u}}}(1-\cos \omega \tau)\right) ,
\label{17}
\end{equation}
with $\tau \equiv \tau_2-\tau_1$, $\epsilon \equiv \alpha^2/2-2$, for $\alpha=2$ one has $\epsilon =0$; while $\epsilon=6$ for $\alpha=4$; and
\begin{equation}
	\tilde{S}_{ins,0}^{\tilde{l}}= \frac{1}{4\pi g \beta}  \left(\frac{2u}{L}\right) (2\pi \tau +2\pi \tilde{l}\beta - \tilde{\delta}\beta)^2 \; . 
\label{adeste}
\end{equation}
Integration over $\tau_1$ and $\tau_2$ or equivalently over $\tau=\tau_2-\tau_1$ and $\tau'=(\tau_1+\tau_2)/2$ gives
\begin{equation}
	\frac{Z_1}{Z_0}= \delta e^{-\frac{8\pi L}{\alpha^2 ug \beta}} \beta \frac{ \int_{-\beta}^\beta d\tau \exp \left[-\tilde{S}_{ins,\omega}\right] 
	\sum_{\tilde{l}=-\infty} ^{+\infty} \exp{\left[-\tilde{S}_{ins,0}^{\tilde{l}}\right]}}{\left. \sum_{\tilde{l}=-\infty} ^{+\infty} \exp{\left[-
	\tilde{S}_{ins,0}^{\tilde{l}}\right]}\right|_{\tau=0}} 
\label{tre}
\end{equation}
with
\begin{equation}
	\sum_{\tilde{l}=-\infty} ^{+\infty} \exp{\left[-\tilde{S}_{ins,0}^{\tilde{l}}\right]} = \sqrt{\frac{gL}{2\beta u}} \theta_3 
	\left( \frac{\pi \tau}{\beta}-\frac{\tilde{\delta}}{2}, e^{-\frac{\pi gL}{2\beta u}} \right) \; .
\end{equation}
From Eq.(\ref{5}), in the high-temperature regime, the supercurrent  (more details on the calculations are given in \ref{App1b})
\begin{equation}
	I_J(\chi)=\delta \frac{8e}{\alpha}e^{-\frac{8\pi L}{\alpha^2 ug\beta}} \left(\frac{2\pi}
	{e^\gamma \Lambda \beta}\right)^{\frac{4}{g}} \beta \sin \frac{2\chi}{ \alpha} 
\label{t}
\end{equation} 
exhibits a power law in $\beta$ whose exponent depends on the strength of the interactions and it is unaffected by 
the angle $\tilde{\delta}$ between the two magnetizations. In the opposite limit of low temperatures  the $2\pi$-periodic JC
\begin{equation}
	I_J(\chi)= \delta \frac{8e}{2} e^{-\frac{4}{g}(\gamma+2\ln 2)}  \left( \frac{2\pi u } {\Lambda L} \right)^{\frac{4}{g}} 
	\frac{2L }{\pi u }4^{\frac{2}{g}-1} \eta_2(\tilde{\delta},g)\sin \chi 
\label{tt} 
\end{equation}
shows a dependance on the angle $\tilde{\delta}$ between the two magnetizations through the modulation function
\begin{equation} 
	\eta_2(\tilde{\delta},g)= \frac{\Gamma{\left(\frac{2\pi+\tilde{\delta}}{\pi g}\right)}\Gamma{\left(\frac{2\pi-\tilde{\delta}}{\pi g}\right)}}
	{\Gamma{\left(\frac{4}{g}\right)}}=\eta_2(-\tilde{\delta},g)
\end{equation}
with $\Gamma(x)$ the  Euler-Gamma function. The modulating function has a minimum in $\tilde{\delta}=0$, i.e. the two magnetizations are parallel and maxima in $\tilde{\delta}=\pm \pi$, i.e. the two magnetizations are anti-parallel and exhibits a weak dependance on $g$.

The $4\pi$-periodic JC is given by 
\begin{equation}
	I_J(\chi)= \delta \frac{8e}{4} \left( \frac{2\pi u } {\Lambda L} \right)^{\frac{4}{g}}\frac{gL}{2u} 4^{\frac{2}{g}} \eta_4(\tilde{\delta},g) \sin \frac{\chi}{2}  
\label{ttt}
\end{equation}
where the modulation function $\eta_4(\tilde{\delta},g)$ can be expressed in terms of the Gaussian Hypergeometric function as
\begin{equation}
	\eta_4(\tilde{\delta},g)=\sum_{s=\pm1} \frac{_2\mathcal{F}_1\left(\frac{4}{g}, \frac{5}{g}+\frac{s \tilde{\delta}}{\pi g};1+\frac{5}{g}+
	\frac{s\tilde{\delta}}{\pi g};-1\right)}{5\pi +s \tilde{\delta}}=\eta_4(-\tilde{\delta},g)
\end{equation}
which has a minimum in $\tilde{\delta}=0$ and maxima in $\tilde{\delta}=\pm \pi$.

\section{Conclusions} \label{section4}
In this paper we studied the parity-dependent Josephson current in a S-2DTI-S junction if $L\gg \xi$ taking into account the Coulomb interaction. 
For transparent S-2DTI interfaces no significant corrections arise with respect to the non-interacting case. When a single magnetic 
impurity whose magnetization is collinear with the spin quantization $z$-axis, the Josephson current is only shifted with respect to 
the transparent regime. If the magnetization lies in the $xy$ plane, the current is sinusoidal and is strongly renormalized by the interaction. 
In particular, for a single barrier at the S-2DTI interface, the current is proportional to $\beta^{-\frac{2}{g}}$ in the high temperature regime and to
 $\L^{-\frac{2}{g}}$ in the low temperature regime. The $2\pi$-periodic critical current is more suppressed by 
repulsive interactions with respect to the $4\pi$-periodic. If two impurities are present at the S-2DTI interfaces new power laws are obtained. 
Remarkably in the low temperature regime both the $2\pi$- and $4\pi$-periodic currents exhibit a dependance on the angle between the two magnetizations.

\ack
We would like to thank B. Trauzettel and F. Cr\'epin for useful comments on the manuscript. This work has been supported by the EU FP7 
Programme under Grant Agreement No. 234970-NANOCTM, IP-SIQS, and by MIUR under project PRIN "Collective quantum phenomena: From 
strongly correlated systems to quantum simulators".

\appendix
\section{} \label{App2}
In this appendix we calculate the effective bosonic action $S_{B}^{eff}$ in the single impurity problem. Let's consider the  Hamiltonian $H_{B}$ 
defined in Eq. (\ref{4}) which can be written as
\begin{equation}
	H_{B}= \frac{u}{2} \int_0^L dx \left( \frac{1}{g} (\partial_x \phi)^2 + g (\partial_x \theta)^2 \right)
\end{equation} 
where $\phi$ and $\theta$ are the bosonic modes of the fields $\Phi$ and $\Theta$ and the corresponding lagrangian
\begin{equation}
	L_{B} = \frac{\partial H_{B}}{ \partial \partial_x \theta}  \partial_x \theta -H_{B} = - \frac{1}{2g} \int_0 ^L dx \left(u (\partial_x \phi)^2 + 
	\frac{1}{u} (\partial_\tau \phi)^2 \right).
\end{equation}
If $L_M$ is the lagrangian of a magnetic impurity in $\overline{x}$, the functional integral $ \int \mathcal{D} \phi \exp \left[ \int_0 ^ \beta d\tau \left(L(\tau)+L_M(\tau)\right) \right]$ 
can be simplified by integrating out the degrees of freedom not involved by the lagrangian $L_M$~\cite{KaneFisher, Furusaki}:
\begin{equation}
	\int \mathcal{D} \phi_{\overline{x}} \ \mathcal{D} \lambda   \mathcal{D} \phi  \exp \left[ \int_0 ^ \beta d\tau \left(L_0(\tau)+L_M(\tau)\right)-
	i\lambda(\tau)(\phi_{\overline{x}}(\tau)-\phi(\overline{x},\tau)) \right] 
\label{20}
\end{equation}
where we have introduced the auxiliary fields $\lambda(\tau)$ and $\phi_{\overline{x}}(\tau)$ whose Fourier series are $\lambda(\tau)=
1/\beta \sum_{\omega} \lambda(\omega) e^{-i\omega \tau}$ and $\phi_{\overline{x}}(\tau)=1/\beta \sum_{\omega} \phi_{\overline{x}} (\omega)
 e^{-i\omega \tau}$. Substituting in (\ref{20}), we get
\begin{equation}
	\eqalign{
	 \fl \int \mathcal{D} \phi_{\overline{x}} \int \mathcal{D} \lambda \exp \left[ -\frac{i}{\beta} \sum_{\omega} 
	 \lambda(\omega) \phi_{\overline{x}} (\omega)\right]\int \mathcal{D} \eta_q \exp 	\left[\frac{1}{\beta}\sum_{\omega} \sum_{q>0} \right.\\
 	\left.\left( \frac{1}{4\pi g} \left(uq +\frac{\omega^2}{uq}\right) |\eta_q(\omega)|^2 -  \sqrt{\frac{\pi}{Lq}} \lambda(\omega)  
	\eta_q (- \omega) \cos q\overline{x}\right) \right] 
	}
\end{equation}
with $\eta_q(\tau)=a^\dagger_q(\tau)-a_q(\tau)$, $q=\frac{\pi n}{L}$, $n \in$  $\mathbb{Z}$, which can be integrated over $\eta_q$ and then 
over the auxiliary field $\lambda$, obtaining
\begin{equation}
	\int \mathcal{D} \phi_{\overline{x}} \exp \left[ - \frac{L}{4\pi u g \beta}  \sum_{\omega} \frac{\omega^2}{\sum_{q>0} 
	\frac{\cos^2 q\overline{x}} {1+\frac{u^2}{\omega^2}q^2}} |\phi_{\overline{x}}(\omega)|^2\right].
\end{equation}
As the sum on $q$ can be exactly evaluated for an arbitrary $\overline{x}$
\begin{equation}
	\frac{\pi}{L}\sum_{q>0} \frac{\cos^2 q\overline{x}} {1+\frac{u^2}{\omega^2}q^2} =  
	-\frac{\pi}{2L} + \frac{ \pi \omega}{4u}\left(1+\frac{\cosh \frac{L \omega}{u} \left(1-\frac{2\overline{x}}{L}\right)}{\cosh\frac{L \omega}{u}} \right) \coth \frac{L \omega}{u},
\end{equation}
we derive the effective action $S_{B}^{eff}$ 
\begin{equation}
	S_{B}^{eff}= -\frac{1}{2\pi g\beta} \sum_{\omega} \frac{\omega^2}{  
	-\frac{u}{L} + \frac{  \omega}{2}\left(1+\frac{\cosh \frac{L \omega}{u} \left(1-\frac{2\overline{x}}{L}\right)}
	{\cosh\frac{L \omega}{u}} \right) \coth \frac{L \omega}{u}} |\phi_{\overline{x}}(\omega)|^2,
\end{equation}
which exhibits a gap of the order $u/L$ if $\omega \rightarrow 0$ as a consequence of the finite size $L$ of the junction.

\section{}
In this appendix some details about the calculation of the partition function $Z$ are given.
\subsection{}
\label{App1a} 

We consider the single impurity problem.
\newline If the  adimensional parameter $A=\beta u/L$ is introduced, the action (\ref{16}) takes the form
\begin{equation}
	\tilde{S}_{ins}=\frac{4\pi}{g} \sum_{n>0}^{n_{max}} \frac{1}{\epsilon A+2\pi n \coth \frac{2\pi n}{A}} 
\label{3maggio} 
\end{equation}
where $n_{max}= \frac{\beta \Lambda}{2\pi}$ corresponds to the cut-off frequency $\Lambda$.
In the high temperature limit $A\ll 1$ and  $ \coth (2\pi n/A) \approx 1+\mathcal{O}(\exp[-2\pi n/A])$, one obtains
\begin{equation}
	\tilde{S}_{ins}  \approx \frac{2}{g} \ln  \left(\frac{e^\gamma \beta \Lambda}{2\pi}\right) +\mathcal{O}(A)
\end{equation}
where $\gamma$ is the Euler-Mascheroni constant. In the low temperature limit $A\gg 1$, Eq. (\ref{3maggio}) can be written as 
\begin{equation}
	\tilde{S}_{ins} =\frac{2}{g} \int_0 ^{\frac{L\Lambda}{u}} dx \frac{1}{x\coth x+\epsilon} 
\label{3maggio1}
\end{equation}
with $x=L \omega/u$. If $\epsilon=0$, namely $\alpha=2$, the integral  (\ref{3maggio1}) can be exactly evaluated 
\begin{equation}
	\tilde{S}_{ins} = \frac{2}{g} \int_0 ^{\frac{L\Lambda}{u}} dx \frac{\tanh x}{x} =\frac{2}{g}\ln \left(\frac{L\Lambda}{\pi u}\right) -\frac{2}{g}\psi \left(\frac{1}{2} \right)
\end{equation}
where $\psi$ is the digamma function~\cite{Gradstein}. If $\epsilon=3$, namely $\alpha=4$, the integral (\ref{3maggio1}) can not be exactly evaluated, but one gets
\begin{eqnarray}
	\tilde{S}_{ins}- \frac{2}{g}\left[\int_0 ^{\frac{L\Lambda}{u}} dx \frac{\tanh x}{x} \right]&=& \frac{2}{g} \left[\int_0 ^{+\infty} \left(-\frac{\tanh x}{x}+ 
	\frac{1}{x\coth x+3}\right) \right]  \nonumber \\
	& \approx& \frac{-3.94}{g}
\end{eqnarray}
and the second integral can be numerically solved. Finally one gets: $2g^{-1}\ln [L\Lambda/(\pi u)] -2g^{-1}\psi(1/2)-3.94g^{-1} \approx 2g^{-1}\ln [L\Lambda/(\pi u)]$, 
because $2\psi(1/2) =-2\gamma - 4\ln 2 \approx -3.92$. The calculation of the partition function follows straightforwardly from (\ref{13}) 
because $\tilde{S}_{ins}$ does not depend on $\tau_1$.

\subsection{} \label{App1b}
We focus on the double impurity problem.
\newline 
The total action $\tilde{S}_{ins}$ has two contributions given by Eqs. (\ref{17}, \ref{adeste}). Firstly we 
consider $\tilde{S}_{ins,\omega}$, namely Eq. (\ref{17}), which can be written as $\tilde{S}_{ins,\omega}\equiv \tilde{S}_{ins1,\omega}+\tilde{S}_{ins2,\omega}$. 
In terms of the adimensional parameter $A$ one has
\begin{equation}
	\tilde{S}_{ins1,\omega} = \frac{8\pi}{g} \sum_{n>0}^{n_{max}} \frac{1}{\epsilon A + 2\pi n \coth \frac{\pi n}{A}} \label{5maggio}
\end{equation}
\begin{equation}
	\tilde{S}_{ins2,\omega} =\frac{4\pi}{g} \sum_{n>0}^{n_{max}}\frac{ \frac{\epsilon A}{2\pi n} + \frac{2}{\sinh \frac{2\pi n}{A}}}{2\pi n + 
	\epsilon A \tanh \frac{\pi n}{A}} \left(1-\cos \frac{2\pi n \tau}{\beta}\right)  
\label{5maggio1}
\end{equation}
In the high temperature limit $A\ll 1$ the dominant contribution to $\tilde{S}_{ins,\omega}$ is given by Eq. (\ref{5maggio}) 
\begin{equation}
	\tilde{S}_{ins1,\omega}  \approx \frac{4}{g}\ln\left(  \frac{e^\gamma \beta \Lambda}{2\pi}\right)  +\mathcal{O}(A), 
\end{equation}
while $\tilde{S}_{ins2,\omega} \propto  \mathcal{O}(A)$. Moreover $\sum_{\tilde{l}=-\infty} ^{+\infty} \exp{\left[-\tilde{S}_{ins,0}^{\tilde{l}}\right]} 
\approx \sqrt{g/A}$ is indipendent on $\tau$ and $\tilde{\delta}$. In the low temperature limit $A\gg 1$, $\tilde{S}_{ins1,\omega}$ can be exactly 
evaluated as in the single barrier case if $\epsilon=0$ and one obtains $\tilde{S}_{ins1,\omega}=4g^{-1}\ln [L\Lambda/(2\pi u)] -4g^{-1}\psi(1/2)$, 
while if $\epsilon=6$ one has the approximated solution $\tilde{S}_{ins1,\omega}=4g^{-1}\ln [L\Lambda/(2\pi u)]$.
 
Let us now consider the contributions to the partition function arising from $\tilde{S}_{ins2,\omega}$. In this case one has to evaluate the integral (\ref{tre}) which can be cast in the form
\begin{equation}
	e^{\frac{\beta u \tilde{\delta}^2}{2\pi gL} } \int_{-\frac{\beta}{2}}^{\frac{\beta}{2}} d\tau e^{-\frac{\beta u}{2\pi gL} \left(\frac{2\pi \tau}{\beta} -
	\tilde{\delta}\right)^2} \exp{\left[-\tilde{S}_{ins2,\omega}\right]}
\label{15maggio}
\end{equation}
because $\sum_{\tilde{l}=-\infty} ^{+\infty} \exp{\left[-\tilde{S}_{ins,0}^{\tilde{l}}\right]} \approx \exp{\left[-\frac{\beta u}{2\pi gL} \left(\frac{2\pi \tau}{\beta} -\tilde{\delta}\right)^2\right]}$.

If $\epsilon=0$, the action $\tilde{S}_{ins2,\omega}$ can be analytically evaluated~\cite{Gradstein} as
\begin{equation}
	\tilde{S}_{ins2,\omega}=\frac{4}{g}\int_0^{+\infty} \;dx\ \frac{1-\cos \frac{xu\tau}{L}}{x\sinh x}=\frac{4}{g} \ln \cosh \left(\frac{\pi  u\tau}{L}\right)
\label{14maggio}
\end{equation}
and from Eq. (\ref{15maggio}) with $x=\pi u \tau/L$ one finally gets 
\begin{equation}
	\frac{2L}{\pi u}  \int_{0}^{+\infty} dx  \cosh \left( \frac{2x\tilde{\delta} }{\pi g} \right) \cosh^{-\frac{4}{g}}x  = 
	\frac{2L}{\pi u } 4^{\frac{2}{g}-1} \frac{\Gamma{\left(\frac{2\pi+\tilde{\delta}}{\pi g}\right)}
	\Gamma{\left(\frac{2\pi-\tilde{\delta}}{\pi g}\right)}}{\Gamma{\left(\frac{4}{g}\right)}}.
\end{equation}
If $\epsilon=6$, $\tilde{S}_{ins2,\omega}$ cannot be exactly evaluated. However, one can see from Eq. (\ref{17}) that dominant corrections to
 $\tilde{S}_{ins2,\omega}$ with respect to the case $\epsilon=0$, are given by
\begin{equation}
	\frac{2\epsilon u}{Lg} \int_0^{+\infty} \;d\omega\ \frac{1- \cos \omega \tau}{\omega^2}= \frac{\epsilon u \pi |\tau|}{Lg} 
\end{equation}
and from Eq. (\ref{15maggio}) with $x=\pi u \tau/L$, one obtains
\begin{eqnarray}
	\frac{2L}{\pi u} \int_{0}^{+\infty} dx e^{-\frac{6x}{g}} \cosh \left( \frac{2x\tilde{\delta} }{\pi g} \right) \cosh^{-\frac{4}{g}}x =\nonumber\\
	\frac{4^{\frac{2}{g}}gL}{2u}\sum_{s=\pm1} \frac{_2\mathcal{F}_1\left(\frac{4}{g}, \frac{5}{g}+\frac{s \tilde{\delta}}{\pi g};1+
	\frac{5}{g}+\frac{s\tilde{\delta}}{\pi g};-1\right)}{5\pi +s \tilde{\delta}}.
\end{eqnarray}
where $_2\mathcal{F}_1$ is the  Gaussian Hypergeometric function.

\end{document}